RESEARCH FEATURE

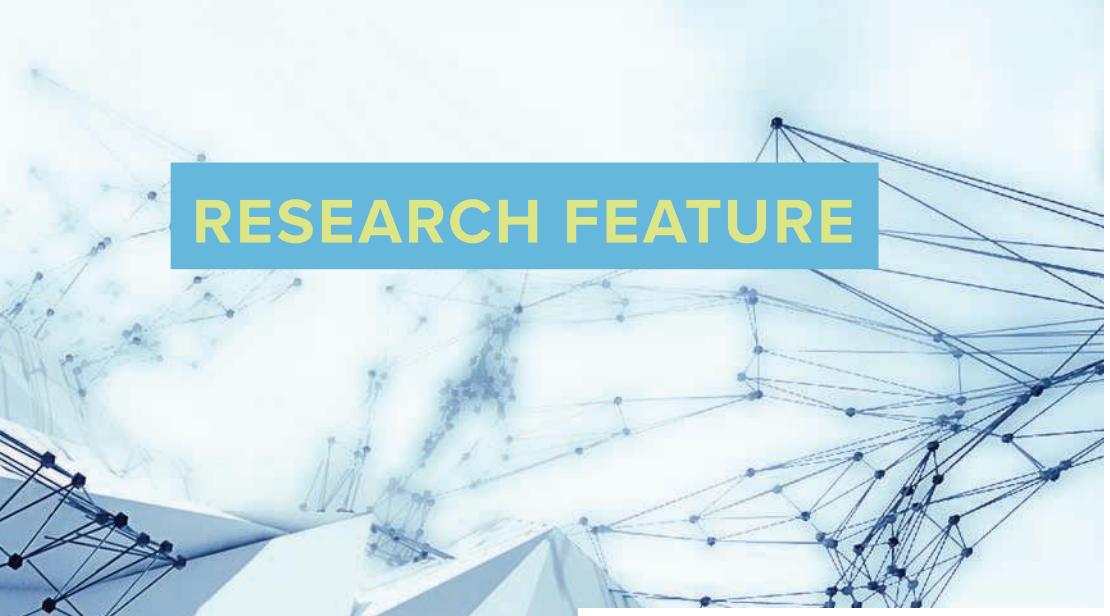
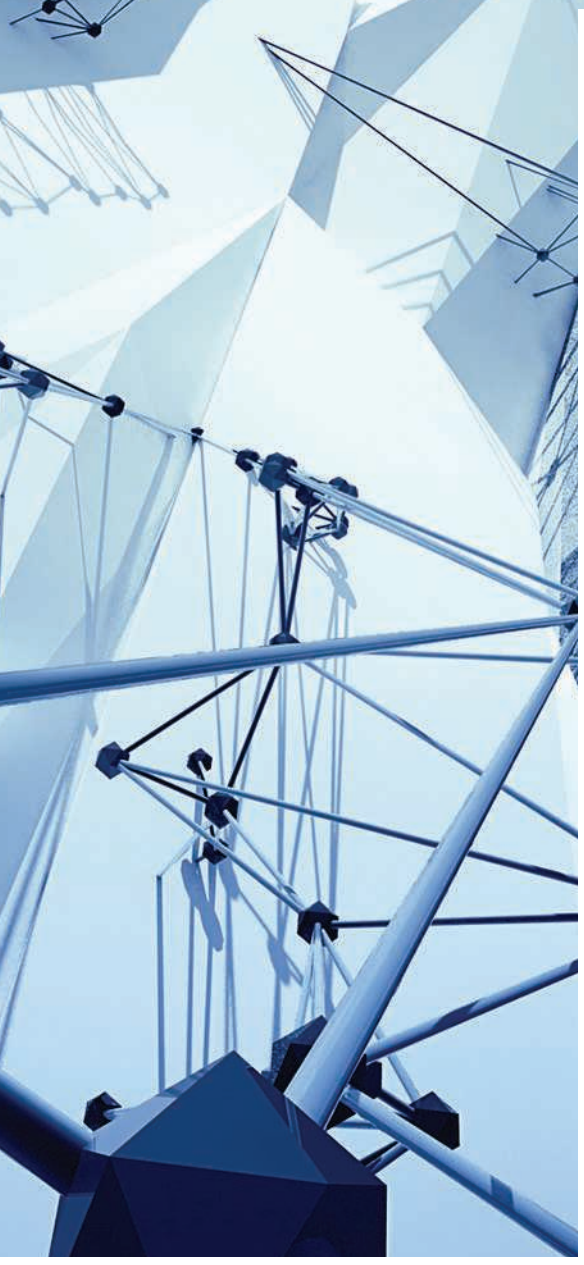

# Decentralized Machine Learning for Intelligent Health-Care Systems on the Computing Continuum

**Dragi Kimovski**, University of Klagenfurt
**Sasko Ristov**, University of Innsbruck
**Radu Prodan**, University of Klagenfurt

*The introduction of electronic personal health records (EHRs) enables nationwide information exchange and curation among different health-care systems. However, current EHR systems are centrally orchestrated, which could potentially lead to a single point of failure.*

Many societies and cultures perceive that sexually transmitted diseases (STDs) only affect "others" who follow "sinful" lifestyles and practices. Therefore, discrimination and stigmatization are common outcomes of such distorted depictions of STDs, especially related to the acquired immunodeficiency syndrome/human immunodeficiency virus. On a global level, the fear of stigmatization prohibits effective disease identification, prevention, care, and treatment adherence, negatively influencing many communities' quality of life.[1]

The introduction of electronic personal health record (EHR) systems is a first step toward addressing these issues, especially for illness-related stigmatization. The purpose of EHR systems is to provide information exchange and curation among different national health-care systems, and support diagnosis quality, safety monitoring, or medical research.

Although EHR systems promise substantial benefits for improved care and reduced health-care costs, there are unforeseen difficulties related to privacy and limited diagnosis-related functionalities. The essential barriers that limit the usability and application of the EHR systems are

- utilization of central control and orchestration, which could potentially lead to a single point of failure, exposure of private information, and hindered interoperability
- storage of personal data in the custody of a single institution, hindering data privacy and limiting knowledge extraction and processing
- limited integration with available personal medical Internet of Things (IoT) devices, such as heart rate monitors or blood sugar sensors
- the inability to utilize a highly heterogeneous set of computing resources.

Consequently, EHRs do not allow intelligence to be injected into the process of medical data analysis. This is primarily due to a lack of basic approaches for supporting decentralized management and transparent integration with medical IoT devices.

Recently, the so-called computing continuum[2] that federates cloud services with emerging fog and edge resources, presented a relevant computing alternative for supporting next-generation EHR systems. The computing continuum provides a vast heterogeneity of computational and communication resources, which can allow low-latency communication for fast decision making close to data sources, and substantial computing resources for a complex data analysis. The distributed nature of the computing continuum further embraces the utilization of machine learning (ML) for the creation of intelligent systems and their federation through distributed ledger technologies (DLTs). It therefore promises transparent integration of omnipresent IoT data, coming for various personal medical sensors. These technologies promise to be the next disruptive ones and will eventually enable intelligently controlled health-care systems with better societal involvement. The execution of ML over the computing continuum and integration with DLTs can make an EHR system personalized, enable transparent integration of IoT devices, and urge active participation of patients and medical professionals in the health-care system. Ultimately, it opens the possibility for training ML algorithms for predictive analysis with medical data belonging to one patient and using the trained models to aid another patient's treatment.

We therefore discuss in this article intelligent computational approaches for an anonymous analysis of medical information across the computing continuum by creating a decentralized ML overlay for model training in a DLT network.

To support such a decentralized EHR system, we explore the possibility of

- creating a decentralized ML network with multiparty computations for secure nonproprietary model training
- a cross-patient predictive analysis for therapy assessment and research with data acquired from medical IoT devices
- a transparent orchestration of the EHR system over heterogeneous resources across the computing continuum.

As a proof of concept, we propose a decentralized conceptual EHR system that uses ML models for anonymous predictive analysis and evaluate it on a real-world computing continuum infrastructure.

## RELATED WORK

### DLTs and decentralized ML for health-care applications

Support for future decentralized platforms for medical data analysis with autonomous practices is being researched extensively. In the literature, Kuo and Ohno-Machado propose a cross-institutional, health-care predictive model for quality-improvement initiatives by predicting the risk of readmission of a group of patients using data from multiple institutions.[3] This approach sets the groundwork for developing privacy-preserving ML technology in a DLT. Furthermore, Mettler provides an initial medical data management approach through DLT, empowering patients and fighting counterfeit drugs in the pharmaceutical industry.[4] Recently, a feasibility study presented in the work of Sheller et al. explore the idea of applying federated learning for secure multi-institutional data analysis with multiple local models coordinated by a centralized aggregation server.[5] Although the concept is promising, it still requires a centralized model to gather all the updates prone to failures and centralized decisions.

Roehrs et al. propose a novel DLT-based architecture, OmniPHR, for a distributed and interoperable EHR architecture.[6] The approach allows for a unified viewpoint of the personal medical information between patients and health-care providers. Furthermore, Roehrs et al. describe a prototype implementation of the OmniPHR architectural model and present an evaluation of the scalability of the approach in terms of integrated, production-ready databases with information from 40,000 adult patients.[7]

The industry has also explored using blockchain for private data storage and management. The GemOS system provides a platform for discovering and sharing disparate data tied to unique identifiers, enabling connections of data sources from different systems on a common ledger and creating proofs of existence with verifiable data integrity.[8]

### Decentralized ML across the computing continuum

Limitations in hospital infrastructures' computational capabilities pose serious problems for deploying ML systems in a decentralized manner. Therefore, the computing continuum has recently been considered as a suitable computing infrastructure, capable of meeting the conflicting requirements of EHR systems.

More concretely, Wang et al. introduce a novel gradient-based training concept for distributed ML models without external computing services over multiple edge resources.[9] The edge devices train local models with local data coming from multiple IoT and medical devices, which are finally aggregated on one device. The work of Kumar et al. presents a novel tree-based ML algorithm called *Bonsai*, for efficient prediction on edge and fog devices close to the IoT devices, which maintains acceptable prediction accuracy while minimizing model size and prediction costs.[10] Furthermore, Osia et al. present a distributed learning approach that complements the cloud for providing privacy-aware and efficient analytics.[11] The algorithm divides the deep learning model into multiple smaller ones, which can be placed on available edge devices while maintaining a central part of the model in the cloud. Moreover, Wang et al. present the application of a deep learning algorithm for medical image analysis that uses fixed-point arithmetic, which can fine-tune the analytic algorithm based on a medical image segment and the available computing resources on the device.[12]

## TECHNOLOGICAL GAPS

### Gap 1: Centralized control of medical data and ML models

Based on the related work analysis, we identified three research technological gaps. Due to privacy concerns, medical institutions manage their data locally, which often leads to inefficient data propagation. This hinders the possibility of training ML algorithms for predictive analysis with medical data belonging to one patient and using the trained models to aid another patient's treatment. The current attempts to integrate ML and IoT with EHR systems, such as federated learning, which enables training across multiple geographically devices, has already yielded promising results.[13] The role of federated learning for EHR systems is twofold: 1) it allows distribution of the training over the computing continuum resources and 2) it brings privacy in combination with the multiparty computing approach. However, even though these algorithms are distributed, they are centrally controlled and require maintenance of a centralized model, periodically updated by multiple local algorithms and potentially exposing private information. In such an environment, malicious federated learning actors can compromise the ML model by mimicking a local or contributing learner/model's role. Even worse, the central actor gathering the model updates (such as those used by Google or Amazon) may steer them toward his or her own personal interests, which may be different from those of the contributors, that is, put biases in the model.

To overcome the identified issues, it is essential to research permissioned DLT protocols for federating a set of ML models, with no need for centralized training and later inference, for three key benefits. First, to improve users' control of the models with a secure relation to their data (or training information). Second, to enable control of information ownership, shared further down the federation of ML models through an adaptive state-transition modeling. Third, to enable anonymous sharing of parts of the ML models (set of rules or wights) between ML systems, with similar characteristics, in the overlay for further improvements.

### Gap 2: Constrained predictive data analysis with limited IoT integration

Training decentralized ML models for predictive analysis of anonymous medical information across DLTs also faces serious challenges. One essential issue is to transparently classify one anonymous patient's medical data among various others through a decentralized collaboration of a set of local ML models (with similar characteristics/algorithms), logically located at different medical institutions. In addition, it is difficult to correlate and analyze millions of anonymous, noncontextualized medical records produced by various devices, distributed into different locations with different attributes. In this scenario, it is difficult to determine whether the data comes from different patients (or even different sensors belonging to the same patient), affecting predictive analysis. Furthermore, the feasibility of training decentralized ML models for medical information analysis, research, and its integration with IoT devices has never been explored.

Therefore, it is important to explore decentralized approaches for the federation of ML training with guided analytics. The approach should address the problem of noncontextualized training data aggregation, knowledge extraction, and cognitive learning about users' medical and personal data in an anonymous manner. This could occur through a seamless coupling of ML predictive analysis algorithms on noncontextualized and anonymous medical information.

### Gap 3: Insufficient computing resources and computationally inefficient DLT and ML solutions for edge training

DLT and ML approaches are known to be computationally demanding.[14] However, in large-scale heterogeneous and fragmented environments where patient data span geographical boundaries, the important limiting factors are insufficient computational resources and technical expertise. Concretely, hospitals do not own a vast infrastructure, and the utilization of high-performance computing services is not always feasible. Furthermore, the employment of local hospital infrastructure for decentralized ML training can lead to reduced accuracy of the ML model and errors during predictive analysis, especially if the medical data for training are generated by IoT devices.

Therefore, we should address scalable approaches for efficient model updates in an ML overlay with an increasing number of learners/algorithms distributed across various physical locations. In practice, we should approach scalability, concerning the available resources across the computing continuum, from various angles such as latency for a consensus and transaction validation time (for example, a model update). It is therefore essential to explore whether we can sacrifice ML model accuracy to allow for execution on computing continuum resources connected directly to personal medical IoT devices (such as heart rate or blood saturation monitors) or other medical equipment, which might not be directly accessible over the network.

## DECENTRALIZED EHR SYSTEM ARCHITECTURE

We propose a conceptual EHR system architecture, named *STIGMA* (see Figure 1). With the STIGMA system, medical data always stay at medical institutions and form local ML models, but only after performing anonymization. Medical professionals interact with the system through multimodal diagnosis equipment, enriched with sensor data from personal IoT devices. Medical institutions can register in the STIGMA EHR system by utilizing strictly defined protocols for interoperability, as defined by Roehrs et al.[6] The STIGMA EHR system performs in the following manner:

- A data analysis instance receives a direct multimodal data stream (magnetic resonance imaging, computed tomography scans, IoT heart rate sensors, electroencephalogram sensors, and so on) of medical procedures.
- Afterward, the data stream is analyzed on the available computing continuum resources.
- Data analysis filters anonymize the data stream, which is then sent to the model training instance.

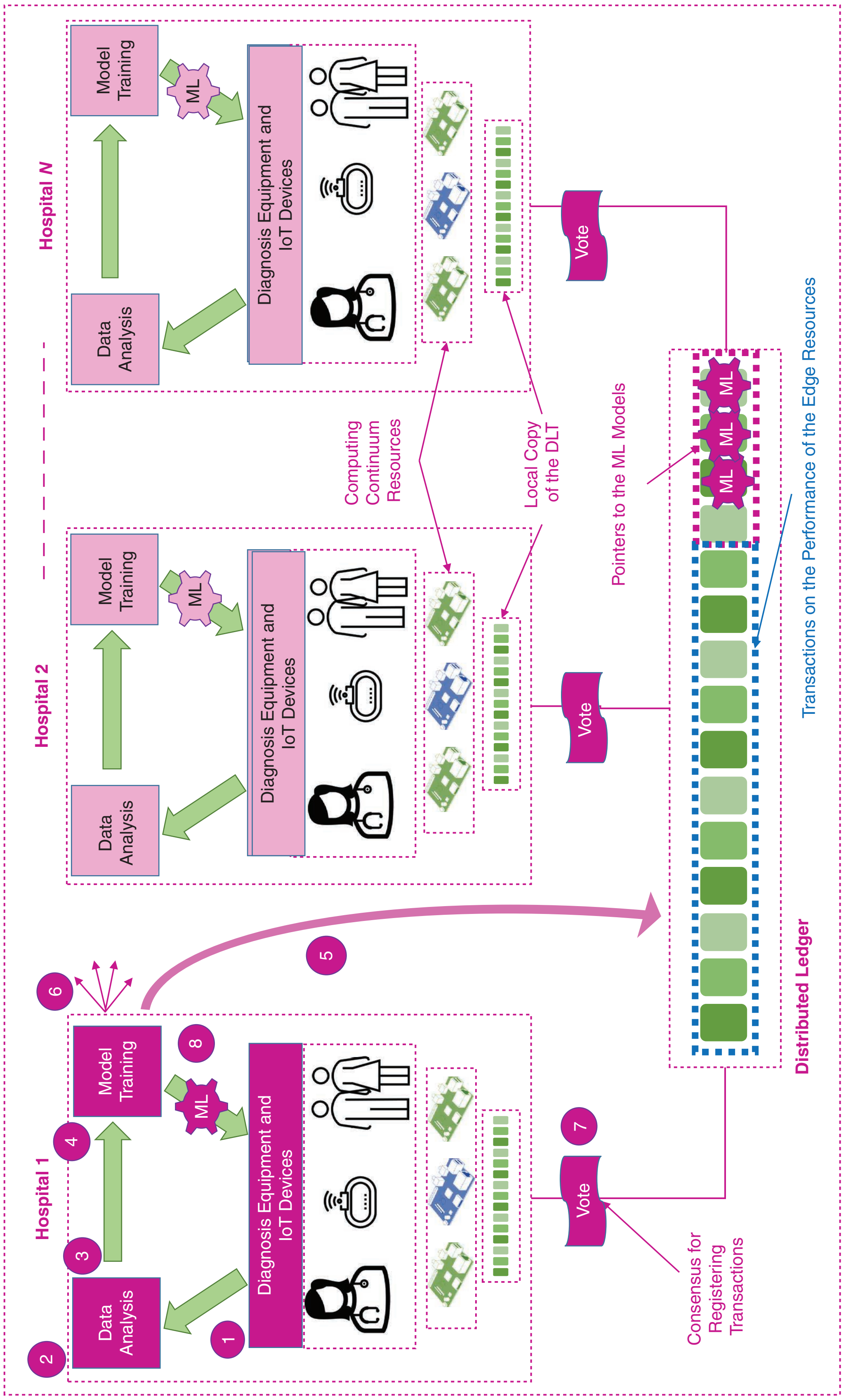

**FIGURE 1.** A conceptual architecture of the STIGMA EHR system.

- The model training instance applies ML algorithms to train a model on the available computing continuum resources.
- After the model is trained, the model training instance utilizes the distributed ledger to register the model (only as a pointer, without exposing the data) and checks for other suitable registered models.
- Thereafter, if suitable models are found in the distributed ledger registry, model training contacts the model owners directly, namely, other medical institutions, to receive rolling updates or exchange (share) relevant data for model improvement.
- The STIGMA EHR system can perform only the rolling updates and the data sharing after a consensus (by voting) is reached among all the medical institutions, federated by the distributed ledger.
- The information is then used for improving the model, which is used to provide real-time support for diagnosis and therapy assessment and is again registered in the distributed ledger.

All of the aforementioned steps are continuously managed and synchronized in a decentralized manner by the STIGMA EHR network. It logically forms a peer-to-peer group that maintains records on all the transactions (model updates, inference performance data, and accuracy). The STIGMA EHR network also contains information for available computing continuum resources (in terms of computing power and available ML models) at each medical institution. The EHR network, therefore, allows all parties of the system to confirm or reject any piece of data added to it, while no data can be deleted from it. This provides a full history of all transactions appearing on the DLT, giving EHRs a method to ensure the correctness of retrieved information.

## ML overlays with decentralized medical data control

To support the creation of a decentralized STIGMA EHR, as depicted in Figure 1, we research a DLT-based overlay for the federation of multiple medical institutions through the following actions, directly related to identified technological gaps.

**DLT for a decentralized federation of ML models in an overlay.** The STIGMA EHR system uses a permissioned[15] protocol to create an appropriate configurable and modular federating architecture addressing EHR systems' requirements for anonymous ML model updates with full control of the private data that do not leave hospital infrastructure. The EHR system relies on scalable DLT management approaches capable of reaching a consensus with a minimal number of communication steps with a limited number of ledgers in a permissioned environment.

**Model provenance for decentralized ML.** Another important aspect of the STIGMA EHR is data provenance, a key concept for supporting ML-based analysis over decentralized networks, especially when data from IoT devices are used. Data provenance enables efficient access approaches that allow all the participating ML models in the ML overlay to maintain a copy of the DLT and ensure the availability of the same version of truth. The DLT contains only the transaction logs that the ML model updates' fingerprints, exclusively stored in hospital computing infrastructures to reduce replication and network throughput.

**Data immutability and secure propagation of decentralized ML model updates with multiparty computation.** This aspect addresses the immutability and propagation of ML models without violating data privacy during updates. It is used to publish, update, and activate anonymous information exchange among the ML algorithms during model training across the overlay. Unfortunately, current technologies are computationally inefficient, thus not allowing straightforward utilization of DLTs for complex data sets. We therefore utilize the concept of multiparty computation[16] to enable computation on data from different providers. The other participating actors gain no additional information about each other's inputs, except what they learn from the ML model's collaborative output, that is, decoupling the model from the training data.

## Predictive data analysis with IoT integration

Another important enabler for deployment of the STIGMA EHR system is cross-medical data analysis for improved diagnosis and therapy assessment through distributed ML with IoT medical device integration. Therefore, we rely on the following solutions.

**Predictive analysis with decentralized ML.** The STIGMA EHR system utilizes approaches for automated ML reasoning with distributed non- and cross-referenced data received from professional medical equipment and personal medical IoT devices. This process reduces uncertainty and mistrust by double-checking the validity of the information and its sources in potentially unpredictable environments. The approach enables shared knowledge and improves data acquisition from IoT devices.

### Scalable ML, and a consensus on the computing continuum

Multiple research works, such as Paxos and RAFT,[17] agree on a single majority value (that is, a state transaction), with reduced overhead and power requirements. Unfortunately, they still require the large computational resources that a resource-constrained hospital infrastructure cannot provide, thus making deployment of the STIGMA EHR system challenging. Therefore, we modify the current approaches to make them suitable for execution on computing continuum devices. To achieve this, the STIGMA EHR system assesses the complexity of ML algorithms and the training data structure to select suitable resources in the computing continuum with higher computational capabilities, close to where the data reside in terms of network distance. Then, based on the available hospital computational infrastructure, a decision is made about where to conduct the training, and the accuracy level is identified.

## REAL TESTBED EVALUATION

To validate the proposed conceptual EHR system, we deployed DLT-based ML systems on a real-world experimental testbed. We emulated the computing infrastructure of medical institutions by using adequate cloud, fog, and edge resources, as described in the "Physical Testbed" section. For the evaluation, we implemented the Paxos three-phase commit protocol, where each institution in the DLT network kept track of current changes. To allow for execution on multiple heterogeneous systems, we developed a Paxos protocol in Java 11.0.

### Physical testbed

We utilized Carinthian Computing Continuum ($C^3$),[18] a real computing continuum testbed, located at the University of Klagenfurt, to emulate a network of multiple medical institutions with limited computing capacities. $C^3$ encompasses heterogeneous resources, provided as containers or virtual machines, in multiple performance categories. We have therefore identified a subset of resources usually available in hospitals (such as fog and private cloud infrastructures), and user-specific devices (such as ECs composed of low-powered, portable devices), to conduct the conceptual evaluation (see Table 1).

Centralized computing infrastructures (CCIs) consist of virtualized instances provisioned on demand from Amazon Web Services. For evaluation purposes, we selected `m5a.xlarge` and `c5.large` as general-purpose instances powered by an AMD EPYC 7000 processor at 2.5 GHz and an Intel Xeon Platinum 8000 series processor at 3.6 GHz, respectively.

A fog cluster (FC) comprises resources from the local Exoscale (ES) cloud provider, which enables communication latency of ≤12 ms and a maximal bandwidth of ≤10 Gb/s. For evaluation purposes, we identified `medium` and `large` ES instances, as described in Table 1.

An edge cluster (EC) includes five NVIDIA Jetson Nano (NJN) and 32 Raspberry Pi-4 (RPi4) single-board computers. We installed a `Raspberry Pi operating system` (version 2020-05-27) on the RPi devices. We used `Linux for Tegra` for the NJN resources. We utilized a

**TABLE 1.** The $C^3$ testbed configuration.

| | CCI (Amazon Web Services) | | FC | | EC | | |
|---|---|---|---|---|---|---|---|
| **Instance/device** | `m5a.xlarge` | `c5.large` | **Exoscale** `large` | **Exoscale** `medium` | **EGS** | **NJN** | **RPi4** |
| CPU type | AMD EPYC 7000 | Intel Xeon Platinum 8180 | Intel Xeon Platinum 8180 | Intel Xeon Platinum 8180 | AMD Ryzen 2920 | Tegra X1 and ARM Cortex A57 | ARM Cortex 72 |
| CPU clock (GHz) | 2.5 | 3.6 | 3.6 | 3.6 | 3.5 | 1.43 | 1.5 |
| Memory (GB) | 32 | 8 | 8 | 4 | 32 | 4 | 4 |
| Storage (GB) | 120 | 120 | 120 | 120 | 1,000 | 64 | 64 |
| BW (Mb/s) | 27 | 26 | 65 | 65 | 813 | 450 | 800 |

FC: fog cluster; EC: edge cluster; EGS: Edge Gateway System; NJN: NVIDIA Jetson Nano; RPi4: Raspberry Pi-4.

managed, 48-port, three-layer HP Aruba switch to interconnect all resources in the EC. The switch supports 1 Gb/s per port, with a latency of 3.8 µs and an aggregate data transfer rate of 104 Gb/s. The EC is managed by the Edge Gateway System (EGS), based on a 12-core AMD Ryzen Threadripper 2920X processor at 3.5 GHz with 32 GB of random-access memory, which is easily available in many medical and business environments. For cases when there are not sufficient resources available at the EC, the EGS is responsible for partially offloading the execution of the compute processes to other computing continuum resources, including FC or CCI.

## Experimental design

We designed the following four sets of experiments according to characteristics of the conceptual decentralized EHR system and averaged the results over 10 runs for statistical significance:

1. The DLT network initialization time evaluates the initialization time of the EHR network, encompassing multiple medical institutions in the range of {3, 5, 7, 10}. The medical institution that initializes the EHR network is considered the first leader, where the leader interval is 30 ms and the delay between voting rounds is 100 ms. The medical institutions join the EHR network in regular intervals of 10 s.
2. The consensus time evaluates the time needed for the network encompassing all medical institutions in the range of {3, 5, 7, 10} to reach a consensus on a single value. Similar to the previous experiment, the leader interval is 30 ms and the delay between voting rounds is 100 ms. The consensus time is measured only after the network is fully initialized with all participating institutions.
3. The ML training time evaluates the training process of a convolutional neural network for object detection with medical multimodal data from laparoscopic procedures[19] limited to 500 samples. The convolutional network has three layers, with a kernel size in the range of {32, 64, 128} and an accuracy of 97%. The ML training time also included the overhead required for transferring the trained model to the device where inference will be performed.
4. Edge accuracy evaluates the tradeoff between accuracy and training time for the aforementioned convolutional neural network on the computing continuum devices. This experiment compares the execution time for training the neural network with an average accuracy of 85 and 70%, respectively.
5. The data transfer time measures the time needed for transfer of 1 MB of raw data between an IoT device, connected to the $C^3$ infrastructure, and the corresponding destination resource. The transfer time was measured using the Prometheus monitoring system.

## Results

Figure 2(a) shows that current consensus algorithms have limited scalability, considering network initialization. We observe that initialization of the EHR network with 10 medical institutions can take up to 28 times more time compared to the small network of three institutions, which limits the number of participating institutions in a single decentralized EHR system. However, the standard deviation ranges from 29% for 10 participating institutions to 58% for three. The reason for the scalability limitation is that all consensus messages must be relayed through a single coordinator, which, although not a single point of failure, is a potential performance bottleneck. This is evident during network initialization for a large number of institutions. However, this experiment proves that up to 10 medical institutions can be federated in a single overlay with minimal initialization overhead.

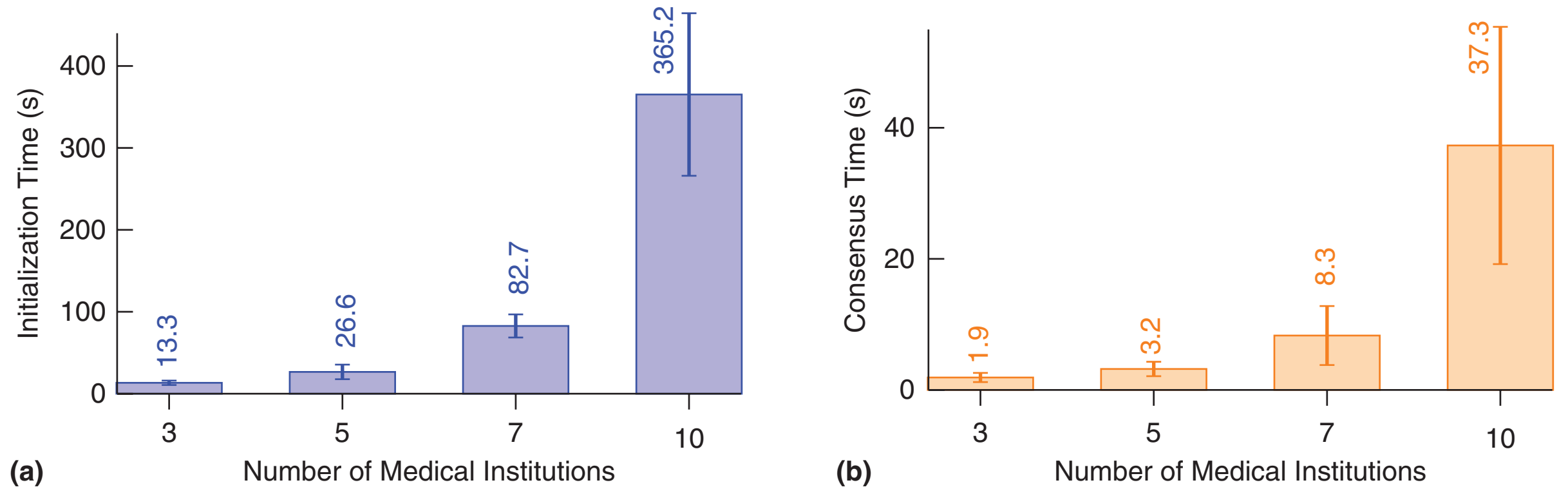

FIGURE 2. A consensus evaluation of the STIGMA EHR system. (a) The DLT network initialization and (b) DLT consensus time.

Furthermore, in Figure 2(b), we observe a similar trend related to the time needed to reach a consensus. The EHR network composed of 10 institutions required nearly 19 times more time to reach a consensus compared to the small network of three institutions. However, we observe a much lower standard deviation, which ranges from 18% for seven participating institutions to 31% for three. Furthermore, compared to the proof-of-work approach implemented in the blockchain protocol, our approach is more efficient in terms or computing resources.

Figure 3(a) evaluates the suitability of the most commonly available resources for performing ML training over multimodal medical data. We observe that specialized devices for ML, such as the NJN device, are very suitable for performing these tasks and can be easily afforded by medical institutions. In addition, available EC devices, extended with other resources from the computing continuum, can achieve very low model-training times, making them suitable for supporting decentralized EHR systems, especially in cases when the system utilization is low. The reason for this is that resources across the computing continuum can meet the conflicting requirements of EHR systems (such as close proximity to the data source and high-performance analysis) due to their high heterogeneity.

Figure 3(b) evaluates the relationship between accuracy of the ML model and execution time on the computing continuum. We observe that by reducing the accuracy from 97 to 85%, we can reduce the execution time by more than 60%. Furthermore, by reducing the accuracy to 70%, we can reduce the execution time on the constrained devices by 90%. However, the tradeoff between accuracy and execution time depends on requirements of the EHR system and the specific medical procedure. In general, this allows for various proximity computing techniques to be applied to improve the performance of ML training without any significant accuracy penalty.

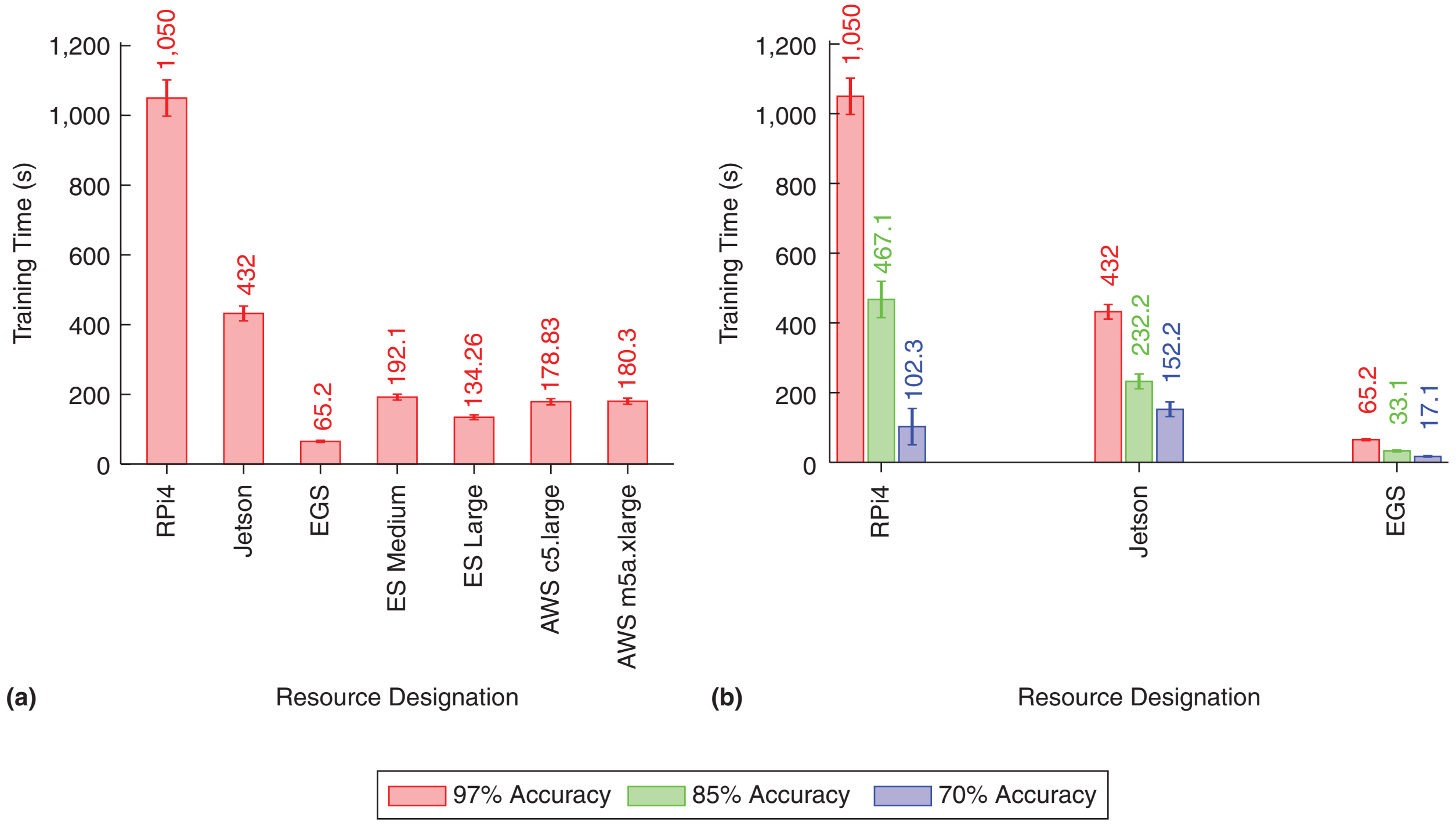

**FIGURE 3.** ML training in the STIGMA EHR system. AWS: Amazon Web Services. (a) The average training time needed to achieve 97% accuracy and (b) average training time needed to lower model accuracy.

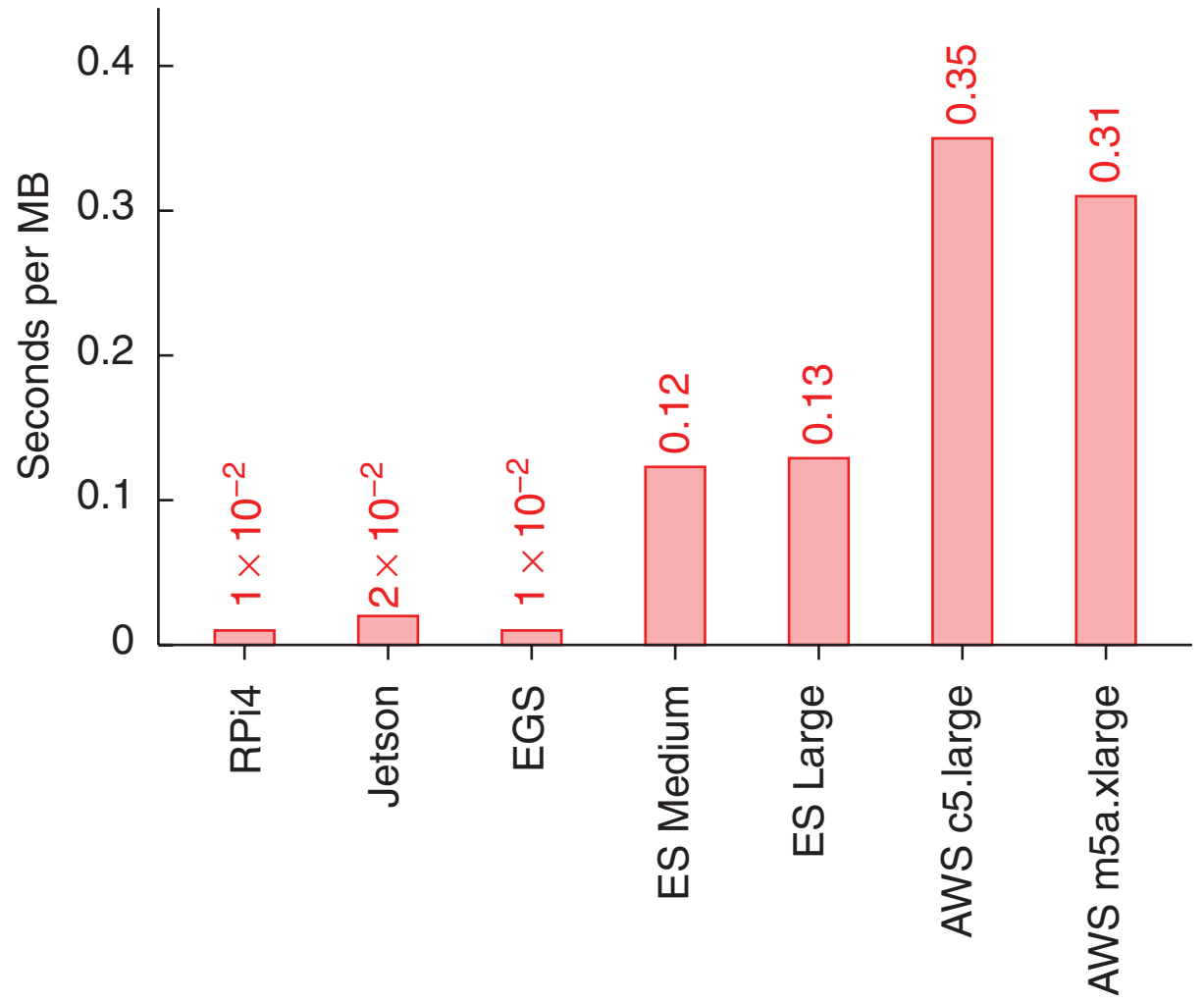

**FIGURE 4.** The effective time for transferring 1 MB of data. AWS: Amazon Web Services.

Finally, Figure 4 analyzes the network performance of raw medical data exchanges among the different resources available across the computing continuum. We observe that the RPi4 and EGS devices can each achieve very low data transfer times compared to the CCI and FC instances, which could significantly reduce any computing performance advantage the CCI alone can provide.

To unleash the potential of decentralized EHR systems and their transparent support by IoT devices, in this article, we explored the need for the creation of an ecosystem that supports the complete lifecycle of medical data sharing and processing. The presented approach enables knowledge extraction for improved medical diagnosis, therapy, and stigma reduction on top of decentralized heterogeneous infrastructures as a part of the computing continuum and IoT environments. We therefore identified critical research gaps. Based on the identified considerations, we defined concrete research and technical actions required for their implementation. Finally, we discussed the implementation of STIGMA, a conceptual, decentralized EHR system as a proof of concept. The system yielded promising results in terms of scalability, which indicate that up to seven different medical institutions can be integrated in a decentralized overlay, with a consensus latency of 8 s or lower. In terms of ML learning time, we observed that edge devices can perform similar to cloud resources, and some of them, such as the EGS, can even reduce training time by 60% compared to the cloud.

Finally, based on the evaluation results of the conceptual STIGMA EHR system, we conclude that decentralized ML over the computing continuum for medical data analysis can be achieved through the utilization of scalable consensus algorithms over a permissioned DLT network with transparent integration of personal IoT devices. In the future, we plan to explore further how we can identify the optimal tradeoff between training accuracy and execution time on low-performance devices across the computing continuum.

## REFERENCES

1. S. E. Stutterheim *et al.*, “Patient and provider perspectives on HIV and HIV-related stigma in Dutch health care settings,” *AIDS Patient Care STDs*, vol. 28, no. 12, pp. 652–665, 2014, doi: 10.1089/apc.2014.0226.
2. P. Beckman *et al.*, “Harnessing the computing continuum for programming our world,” *Fog Comput., Theory Pract.*, pp. 215–230, Apr. 2020, doi: 10.1002/9781119551713.ch7.
3. T.-T. Kuo and L. Ohno-Machado, “Modelchain: Decentralized privacy-preserving healthcare predictive modeling framework on private blockchain networks,” 2018, *arXiv:1802.01746*.
4. M Mettler. “Blockchain technology in healthcare: The revolution starts here,” in *Proc. 2016 IEEE 18th Int. Conf. e-Health Netw., Appl. Services (Healthcom)*, pp. 1–3, doi: 10.1109/HealthCom.2016.7749510.
5. M. J. Sheller, G. Anthony Reina, B. Edwards, J. Martin, and S. Bakas, “Multi-institutional deep learning modeling without sharing patient data: A feasibility study on brain tumor segmentation,” in *Proc. Int.*

## ABOUT THE AUTHORS

**DRAGI KIMOVSKI** is a tenure-track researcher at the Institute of Information Technology, Klagenfurt University, Klagenfurt, 9020, Austria. His research interests include fog and edge computing, multiobjective optimization, and distributed storage. Kimovski received a Ph.D. in computer science from the Technical University of Sofia. Contact him at: dragi.kimovski@aau.at.

**SASKO RISTOV** is a postdoctoral university assistant at the University of Innsbruck, Innsbruck, 6020, Austria. His research interests include performance modeling and optimization of parallel and distributed systems, particularly workflow applications and serverless computing. Ristov received a Ph.D. in computer science from The Saints Cyril and Methodius University of Skopje. Contact him at sashko@dps.uibk.ac.at.

**RADU PRODAN** is a professor of distributed systems at the Institute of Information Technology, Klagenfurt University, Klagenfurt, 9020, Austria. His research interests include performance and resource management tools for parallel and distributed systems. Prodan received a Ph.D. in computer science from the Vienna University of Technology. Contact him at radu.prodan@aau.at.

*MICCAI Brainlesion Workshop,* 2018, pp. 92–104.
6. A. Roehrs, C. A. Da Costa, and R. da Rosa Righi, "OmniPHR: A distributed architecture model to integrate personal health records," *J. Biomed. Inf.*, vol. 71, pp. 70–81, Jul. 2017, doi: 10.1016/j.jbi.2017.05.012.
7. A. Roehrs, C. A. da Costa, R. da Rosa Righi, V. F. da Silva, J. R. Goldim, and D. C. Schmidt, "Analyzing the performance of a blockchain-based personal health record implementation," *J. Biomed. Inf.*, vol. 92, pp. 103,140, Apr. 2019, doi: 10.1016/j.jbi.2019.103140.
8. "The GemOS System," Gem, 2021. https://enterprise.gem.co/gemos/ (Accessed: May 5, 2021).
9. S. Wang *et al.*, "When edge meets learning: Adaptive control for resource-constrained distributed machine learning," in *Proc. IEEE INFOCOM 2018-IEEE Conf. Comput. Commun.*, pp. 63–71, doi: 10.1109/INFOCOM.2018.8486403.
10. A. Kumar, S. Goyal, and M. Varma, "Resource-efficient machine learning in 2 kb ram for the internet of things," in *Proc. 34th Int. Conf. Mach. Learn.*, 2017, vol. 70, pp. 1935–1944.
11. S. A. Osia, A. S. Shamsabadi, A. Taheri, H. R. Rabiee, and H. Haddadi, "Private and scalable personal data analytics using hybrid edge-to-cloud deep learning," *Computer*, vol. 51, no. 5, pp. 42–49, 2018, doi: 10.1109/MC.2018.2381113.
12. G. Wang *et al.*, "Interactive medical image segmentation using deep learning with image-specific fine tuning," *IEEE Trans. Med. Imag.*, vol. 37, no. 7, pp. 1562–1573, 2018, doi: 10.1109/TMI.2018.2791721.
13. T. S. Brisimi, R. Chen, T. Mela, A. Olshevsky, I. C. Paschalidis, and W. Shi, "Federated learning of predictive models from federated electronic health records," *Int. J. Med. Inf.*, vol. 112, pp. 59–67, Apr. 2018, doi: 10.1016/j.ijmedinf.2018.01.007.
14. T.-T. Kuo, H.-E. Kim, and L. Ohno-Machado, "Blockchain distributed ledger technologies for biomedical and health care applications," *J. Amer. Med. Inf. Assoc.*, vol. 24, no. 6, pp. 1211–1220, 2017, doi: 10.1093/jamia/ocx068.
15. E. Gaetani, L. Aniello, R. Baldoni, F. Lombardi, A. Margheri, and V. Sassone, "Blockchain-based database to ensure data integrity in cloud computing environments," in *Proc. Italian Conf. Cybersecurity,* 2017, pp. 1–10.
16. O. Goldreich, "Secure multi-party computation," Weizmann Inst. Sci., Rehovot, Israel, 1998. [Online]. Available: https://citeseerx.ist.psu.edu/viewdoc/download?doi=10.1.1.11.2201&rep=rep1&type=pdf
17. D. Ongaro and J. K. Ousterhout, "In search of an understandable consensus algorithm," in *Proc. USENIX Annu. Tech. Conf.*, 2014, pp. 305–319.
18. D. Kimovski, R. Mathá, J. Hammer, N. Mehran, H. Hellwagner, and R. Prodan, "Cloud, fog or edge: Where to compute?" *IEEE Internet Comput.*, vol. 25, no. 4, pp. 30–36, 2021, doi: 10.1109/MIC.2021.3050613.
19. A. Leibetseder, S. Kletz, K. Schoeffmann, S. Keckstein, and J. Keckstein, "GLENDA: Gynecologic laparoscopy endometriosis dataset," in *Proc. 26th Int. Conf., MultiMedia Modeling (MMM)*, Daejeon, South Korea, Jan. 5–8, 2020, pp. 439–450, doi: 10.1007/978-3-030-37734-2_36.